\documentclass[journal]{IEEEtran} 
\IEEEoverridecommandlockouts
\usepackage{cite}
\usepackage{amsmath,amssymb,amsfonts}
\usepackage{algorithmic}
\usepackage{graphicx}
\usepackage{textcomp}
\usepackage{xcolor}
\usepackage{subcaption}
\usepackage{physics}
\usepackage{upgreek}
\usepackage{multirow}

\usepackage{flafter}

\def\BibTeX{{\rm B\kern-.05em{\sc i\kern-.025em b}\kern-.08em
    T\kern-.1667em\lower.7ex\hbox{E}\kern-.125emX}}

\newcommand{\Fig}{Fig.}

\newcommand{\micron}{\(\upmu\)m}

\makeatletter
\newcommand{\linebreakand}{%
  \end{@IEEEauthorhalign}
  \hfill\mbox{}\par
\mbox{}\hfill\begin{@IEEEauthorhalign}
}
\makeatother
    
\begin{document}

\title{A Multiscale Workflow for Thermal Analysis of 3DI Chip Stacks
\thanks{Funding provided by IBM-RPI Future of Computing Research Collaboration (FCRC)}
}
\author{
    \IEEEauthorblockN{Max Bloomfield\textsuperscript{*}}%
    , \IEEEauthorblockN{{Amogh Wasti}\textsuperscript{*}}%
    , \IEEEauthorblockN{{Zongmin Yang}\textsuperscript{*}}%
    , \IEEEauthorblockN{{Matthew Galarza}\textsuperscript{*}}%
    , \IEEEauthorblockN{{Theodorian Borca-Tasciuc}\textsuperscript{*}}%
    , \IEEEauthorblockN{{Jacob Merson}\textsuperscript{*}}%
    , \IEEEauthorblockN{{Timothy Chainer}\textsuperscript{\dag}}%
    , \IEEEauthorblockN{Prabudhya Roy Chowdhury\textsuperscript{\ddag}}%
    , \IEEEauthorblockN{Aakrati Jain\textsuperscript{\ddag}}\\
    \IEEEauthorblockA{{\textsuperscript{*}\textit{Rensselaer Polytechnic Institute, Troy NY, USA}}}\\%
    \IEEEauthorblockA{\textsuperscript{\dag}\textit{T.J. Watson Research Center, Yorktown Heights NY, USA}}\\%
    \IEEEauthorblockA{\textsuperscript{\ddag}\textit{IBM Research, Albany NY, USA}}%
    }

\maketitle

\begin{abstract}
Thermally aware design of 2.5D and 3D advanced packaging systems will require fast, accurate, and powerful thermal analysis of chiplets, stacks, and packages. These systems contain multiple materials with non-linear heat transfer properties and geometric feature sizes that span many orders of magnitude. The smallest heterostructures in the front and back ends of the line present significant thermal modeling and analysis challenges in isolation. Replicated millions or billions of times in a chiplet stack, these structures present a near insurmountable hurdle to meeting the speed and accuracy needed of analysis in the design process. Additionally, establishing precise parameter values for the materials in these systems, when size and temperature dependencies create significant deviations from bulk properties, further complicates the problem.

To address these issues, we have developed a multiscale methodology that advances the current state of the field by enabling die-scale simulations that capture phenomena arising from the structural details of the BEOL metallization stack. Taking advantage of the large length-scale separation between the BEOL features and the die-level structures, we employ a hierarchical, multiscale, finite-element approach. This hierarchical method uses a standard finite element method (FEM) formulation on a die or package scale, using computational homogenization to obtain effective thermal conductivities in the BEOL. Referring to industry-standard layout and design files, we construct and solve a locally appropriate subscale FEM problem in a representative volume element (RVE) at every quadrature point in the macroscale FEM problem.
To accomplish this calculation, in our multiscale workflow, all geometric models of these RVEs are automatically constructed, meshed, and used to compute homogenized, anisotropic, thermal conductivities from the relevant GDSII or OASIS files based on the FEM integration point locations. Here, we make use of a direct, static-condensation based method to extract the full thermal conductivity tensor.

\end{abstract}


\section{Introduction}
As the industry moves toward 2.5D and 3D heterogeneous integration (HI) targets, thermal management has become an increasing concern. Various heat removal strategies exist and continue to be developed, but it is difficult to put them to best use without reliable insights into how heat will distribute inside a proposed stack with incredibly complex thermal conduction paths. Progress has been made towards package-level thermal simulations,  via block homogenization on a variety of scales, from a chiplet or bond scale~\cite{jainThermalCharacterization3D2023,jainThermalPerformanceCharacterization2021,salviReviewRecentResearch2021} to interposers~\cite{ma2014,minghaozhou2022}. However, chiplet- or die-scale solutions which account for the detailed structure of the BEOL have remained computationally infeasible within the time and accuracy constraints of the design process. 
 The current work indicates that the difficulty of efficiently creating the necessary geometric models for such fine-grained analysis of chiplet scale may be less than previously estimated~\cite{pfromm2024mfitmultifidelitythermalmodeling} when a large separation exists between a representative subscale and the scale of the desired resolution of the macroscale. This paper focuses on the development of a new, automated, multiscale workflow that can be used to perform rapid analysis of 3D HI designs that include the in-plane spatial heterogeneity of the BEOL.

\section{Macroscale Model}
Since a multiscale scheme is used to model the BEOL structures, we split our description into two components. The macroscale model, described here, represents the thermal behavior on the chiplet scale. It makes use of a standard finite element method (FEM) formulation for linear heat conduction. In this work, the macroscale model is solved for a temperature distribution using the MUMFiM multiscale framework \cite{mersonNewOpensourceFramework2024}; however, any FEM solver such as ANSYS~\cite{ansys} or Abaqus~\cite{abaqus} could be used. The geometric model and boundary conditions on this package scale model are described here.

\subsection{Geometry}
\begin{figure}
    \centering
    \includegraphics[width=\linewidth]{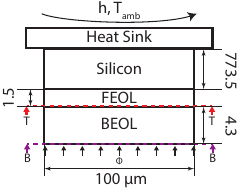}
    \caption{Front view of the proxy-chip modeled in this paper. In the multiscale model, the BEOL properties are extracted from a test vehicle. A heat flux with integrated power of 1.256~mW is applied to the bottom surface. The top surface is exposed to an ambient temperature of 40\textsuperscript{$\circ$} C with a convection coefficient, $h$, of 4.0 W/K mm\textsuperscript{2}. All dimensions are reported in \micron{}. The red dashed line (T-T) and purple dashed line (B-B) represent the top and bottom of the BEOL, respectively, where the temperature fields are reported. In the text, these surfaces are referred to as top surface and bottom surface respectively. }
    \label{fig:model-chip}
\end{figure}

 The chip is modeled as a rectangular prism with a cross-sectional area of 10,000~\micron{}\textsuperscript{2} (100 x 100~\micron{}). Through the thickness it is split into three domains: a silicon layer that is 773.5~\micron{} thick with a thermal conductivity of 139.4 W/m-K; a currently unpowered active device region (FEOL) with a thickness of 1.5~\micron{} and modeled as silicon with a thermal conductivity of 139.4 W/m-K; and a BEOL layer 4.3~\micron{} thick and is modeled using anisotropic conductivity tensor values derived from our multiscale scheme. This macroscale geometry is shown schematically in \Fig{} \ref{fig:model-chip} and the thermal properties are summarized in Table \ref{tab:matl-props}. The heat sink is modeled through a convective boundary condition, but its structure is not explicitly included in the computational domain.

\begin{figure}
    \centering
    \includegraphics[width=\linewidth]{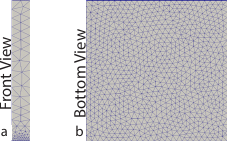}
    \caption{In panel a, the front view of the macroscale model is shown with a highly graded tetrahedral mesh to minimize the total number of elements in regions far away from the thin BEOL and FEOL layers. Panel b shows the mesh on the bottom surface of the chip stack (c.f.,~\Fig{}~\ref{fig:model-chip}, surface B-B).
    }
    \label{fig:mesh}
\end{figure}

The macroscale model contains 12,537 linear tetrahedral finite elements, with 5,105 elements in the BEOL region. The mesh is shown in \Fig{}~\ref{fig:mesh}. To an experienced modeler, this mesh may seem surprisingly coarse for a 100 by 100~\micron{} cross-sectional area when accounting for the smallest features in the BEOL. The small size of this mesh demonstrates the effectiveness of our approach, affording the use of homogenization and fine elements in only the domains where they are required. Here, we have made use of a highly graded mesh to reduce the number of elements far away from the thin BEOL and FEOL structures. Without the homogenization of the subscale structures, capturing their effect on heat transfer via direct representation in the mesh would require a far larger (perhaps infeasible) number of mesh elements for a domain of this size.

\begin{table}[!t]
\caption{Dimensions and materials for the components of the 3D package.}
\centering
\begin{tabular}{|l|l|l|l|}
\hline
\textbf{Component} & \textbf{\begin{tabular}[c]{@{}l@{}}Thickness\\ (micron)\end{tabular}} & \textbf{Material} & \textbf{\begin{tabular}[c]{@{}l@{}}Thermal\\ Conductivity (W/m-K)\end{tabular}} \\ \hline
Silicon            & 773.5                        & Silicon              & 139.4                          \\ \hline
FEOL            & 1.5                         & Silicon              & 139.4                          \\ \hline
BEOL               & 4.3                         & Composite         & \begin{tabular}[c]{@{}l@{}}Homogenized,\\ Spatially Varies\end{tabular}  \\ \hline
\end{tabular}
\label{tab:matl-props}
\end{table}

\subsection{Boundary Conditions}
In this work, we use a set of macroscale boundary conditions that is designed to mimic the topmost chiplet in a 3D multi-chiplet stack. The choice of boundary conditions and model geometry is consistent with having additional heat-dissipating chiplets below the bottom surface of the stack shown in \Fig{}~\ref{fig:model-chip} (labeled B-B). In this work, we restrict our model to passive heat transfer without powering on active devices in the FEOL layer.

\begin{figure}[!t]
    \centering
    \includegraphics[width=\linewidth]{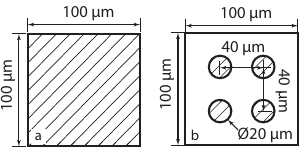}
    \caption{A schematic of the heat flux on the bottom surface (100$\times$100~\micron{}) is shown, where flux is applied to the regions with hash marks. In both cases, a power input flux of 1.256 mW/mm\textsuperscript{2}\ is applied. In the uniform case (a) the heat flux is applied uniformly across the bottom surface. In the non-uniform case (b) the heat flux is applied in circular domains on the bottom surface representing the contact patches of microbumps.}
    \label{fig:heat-flux-application}
\end{figure}

On bottom surface (c.f.,~\Fig{}~\ref{fig:model-chip}, surface B-B), heat flux is prescribed in two loading patterns that are schematically shown in \Fig{}~\ref{fig:heat-flux-application}. In the uniform case (\Fig~\ref{fig:heat-flux-application}a), a heat flux of 0.1256~W/mm\textsuperscript{2} is applied uniformly across the bottom surface. In the non-uniform case (\Fig~\ref{fig:heat-flux-application}b) a heat flux of 1.0~W/mm\textsuperscript{2} is applied in an array of four circular domains on the bottom surface, each with a diameter of 20~\micron{} and a center-to-center distance of 40~\micron{}. These circular domains represent the contact patches of copper microbumps. These nominal conditions were chosen to make the integrated thermal power applied across the bottom surface approximately equal for both cases, with a value of 1.256~mW/mm\textsuperscript{2}.

As shown in \Fig{}~\ref{fig:model-chip}, on the top of the computational domain, a convective boundary condition is applied to represent contact with a heat sink. Here, we choose the ambient temperature \((T_\text{amb})\) to be \(40^\circ\)~C and the convective heat transfer coefficient ($h$) to be a nominal value of 4.0~mW/K mm$^2$. Adiabatic boundary conditions are applied to the vertical sides of the domain.

\section{Multiscale BEOL Model} \label{sec:multiscale-model-formulation}
To capture the spatial fluctuations of the BEOL, we make use of an upscaling scheme that automatically constructs a homogenized thermal conductivity tensor for each integration point of the macroscale finite element mesh. In this scheme, we take advantage of strong scale separation, or the assumption that the subscale behavior can be accounted for by a small geometric domain around each integration point that can be considered ``differential'' compared to the macroscale element.These small regions should have representative material properties of the continuum model in that region and are called representative volume elements (RVEs). 

This assumption of strong scale separation affords a first order Taylor series representation of the temperature field as
\begin{multline}
    T^\text{s} (\vb{X}^{\text{M}},\vb{X}^{\text{s}}) = \\T^\text{M}(\vb{X}^{\text{M}}) + \nabla T^{\text{M}}(\vb{X}^{\text{M}}) \cdot (\vb{X}^{\text{s}}-\vb{X}^{\text{M}}) + \Tilde{T}^{\text{s}} (\vb{X}^{\text{M}},\vb{X}^{\text{s}}),
    \label{eq:micro_temp}
\end{multline}
where, \(T^\text{M}\) is the macroscale temperature, \(T^\text{s}\) is the subscale temperature, \(\vb{X}^\text{M}\) is the macroscale coordinate (centroid of the RVE), \(\vb{X}^\text{s}\) is the subscale coordinate which is referenced to the RVE centroid, and \(\Tilde{T}^{\text{s}}\) is known as the subscale temperature fluctuation. By placing the RVE centroid at the location of the macroscale coordinate, we obtain \(<\vb{X}^{\text{s}} - \vb{X}^{\text{M}} > = \vb{0}\), where \(<\cdot>\) represents the volume average.

\begin{enumerate}
    \item The macroscopic temperature field is equal to the volume average of the subscale temperature field, such that
    \begin{equation}
        T^{\text{M}} = <T^\text{s}>.
        \label{eq:theta_homogenization}
    \end{equation}
    \item The macroscopic temperature gradient field is equal to the volume average of the subscale temperature gradient field, such that
    \begin{equation}
        \nabla T^{\text{M}} = <\nabla T^{\text{s}}>.
        \label{eq:theta_grad_homogenization}
    \end{equation}
    
    \item For a steady-state system, the macroscale heat flux is equal to the volume average of the microscale heat flux, such that
    \begin{equation}
        \vb{q}^{\text{M}} = <\vb{q}^{\text{s}}>.
    \end{equation}
\end{enumerate}

The combination of \eqref{eq:micro_temp} with \eqref{eq:theta_homogenization} and \eqref{eq:theta_grad_homogenization} provides the additional constraints as
\begin{equation}
    <\Tilde{T}^{\text{s}}>=0,
\end{equation}
and
\begin{equation}
    <\nabla \Tilde{T}^{\text{s}}>=0.
\end{equation}
This leads to the boundary condition
\begin{equation}
    \int_{\Gamma^{\text{s}}}\Tilde{T}^{\text{s}} \vb{n} \dd{ \Gamma^{\text{s}}} = \vb{0},
\end{equation}
which is compatible with the extended Hill-Mandel criteria.

In this work, we do not directly compute the subscale temperature field and heat flux, but instead use these rules to develop a homogenized thermal conductivity tensor as described in section \ref{sec:conductivity-extraction}.

\subsection{RVE Generation}
\begin{figure*}
    \centering
    \includegraphics[width=\linewidth]{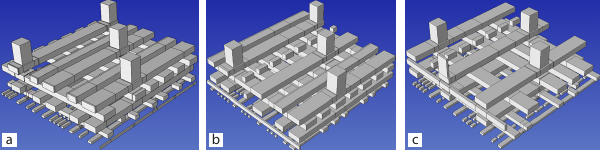}
    \caption{Eleven layers of metal components in exemplar RVEs (five layers of lines and six layers of vias). Dielectric layers have been excluded from the figure so that the metal structures with high conductivity are visible. Dielectrics are included in the multiscale simulation. The RVE in each panel is extracted from different locations in the test vehicle.}
    \label{fig:rve-exemplar}
\end{figure*}

In this work, we construct RVEs that are 2$\times$2~\micron{} and span the full thickness of the BEOL (4.3~\micron{}). The RVEs contain several metallization layers, seen in \Fig{} \ref{fig:rve-exemplar}, which shows three exemplar RVEs. In total, we construct 5,105 independent RVEs. These RVEs are generated through a semi-automatic workflow process that follows the following steps:

\begin{enumerate}
    \item Manually translate process spreadsheet with layer thickness and material properties to XML file.
    \item At each quadrature point in the macroscale mesh, the XML file and GDSII or OASIS files are automatically read into a custom model generation utility that constructs a 3D CAD model.
    \item The generated geometry is automatically read into a custom MuMFiM module that computes the homogenized properties for the RVEs.
\end{enumerate}

The geometric model construction routines use a CAD modeling kernel (Parasolid, version 37.0~\cite{parasolid}) to extrude the contours extracted from the appropriate location and layers in a GDSII layout and perform the constructive solid geometry operations to form the lines, vias, and surrounding dielectric layers. Each material is assigned appropriate thermal conductivities during this construction process, as shown in Table~\ref{tab:matl-props}.

\subsection{Conductivity Extraction} \label{sec:conductivity-extraction}
Once the 3D model is constructed, the thermal conductivity can be computed. In this work we directly compute the thermal conductivity using a static-condensation-based method rather than making use of finite differencing schemes. This method is more accurate and can be effectively extended to temperature-dependent material properties and nonlinear heat conduction problems \cite{waseemModelReductionComputational2020}. In essence, this static condensation method reduces the finite element problem to a new basis that only has degrees of freedom on the RVE boundary.

\subsection{Macroscale Finite Element Model}
On the macroscale, we solve the steady state heat equation summarized as follows,
\begin{equation}
    \nabla \cdot \vb{q} = Q \qquad x \in V
\end{equation}
where,  \(\vb{q}\) is the heat flux, \(Q\) is the volumetric body heat load, and \(V\) is the macroscale volume. We make use of Fourier heat conduction
\begin{equation}
    \vb{q} = -\vb*{\kappa} \nabla T
    \label{eq:fourier}
\end{equation}
where, \(\vb*{\kappa}\) is the thermal conductivity tensor and \(T\) is the temperature field. In our multiscale scheme, the subscale RVE model provides the thermal conductivity tensor at each material point of our discretized finite element model. This type of effective medium approximation for heat transfer has been discussed and validated in work by Wasti, \textit{et al}\cite{SHTC2025}. The thermal conductivity tensor is symmetric and positive definite and can be written as
\begin{equation}
    \vb*{\kappa} = \begin{bmatrix}
        \kappa_{xx} & \kappa_{xy} & \kappa_{xz} \\
         & \kappa_{yy} & \kappa_{yz} \\
        \text{sym.} &   & \kappa_{zz}\\
    \end{bmatrix}.
\end{equation}

This model permits a temperature boundary condition,
\begin{equation}
   T = T_D \qquad \in \Gamma_D,
\end{equation}
flux boundary condition,
\begin{equation}
   -\vb{q} \cdot \vb{n} = \phi \qquad \in \Gamma_N,
\end{equation}
and, convection boundary condition,
\begin{align}
   \vb{q} \cdot \vb{n} = h (T-T_\text{amb}) \in \Gamma_R,
\end{align}
where, \(\Gamma=\Gamma_D\cup\Gamma_N\cup\Gamma_R\) and \(\Gamma_D\cap\Gamma_N\cap\Gamma_R = \emptyset \).

In this work, we solve the steady state heat equation weakly making use of the finite element method. The weak problem is stated as: Given \(T_D\), \(T_{\text{amb}}\), \(h\), \(\vb*{\kappa}\), \(Q\), \(\phi\) find \(T \in \delta\) such that \(\forall w \in \mathcal{V} \):
\begin{multline}
    \int_V \nabla w \cdot \vb*{\kappa} \cdot \nabla T \dd{V} + \int_{\Gamma_R} w h T \dd{\Gamma} = \\ \int_V w Q \dd{V} + \int_{\Gamma_N} w \phi \dd{\Gamma} +\int_{\Gamma_R} w h T_\text{amb} \dd{\Gamma},
\label{eq:weak-form}
\end{multline}
where,
\begin{align}
    &\delta = \{T\ |\ T \in H^1,\ T|_{\Gamma_D}=T_D\} \\
    &\mathcal{V} = \{w\ |\ w \in H^1,\ w|_{\Gamma_D}=0\},
\end{align}
and, \(H^1\) represents the standard Sobolev space including square integrable first derivatives.

In this work, we use a standard Galerkin finite element implementation that is, we use the same shape functions for the test functions (\(w\)) and trial functions (\(T\)).




\section{Results and Discussion}

\begin{figure}
    \centering
    \includegraphics[width=\linewidth]{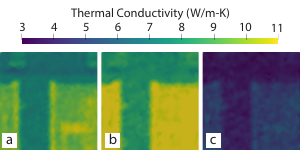}
    \caption{Homogenized BEOL thermal conductivities in the 100$\times$100~\micron{} cross-section. We see large spatial heterogeneity that will impact chip-scale hotspot locations. The in-plane components are shown in panel a (\(\kappa_{xx}\)) and b (\(\kappa_{yy}\)). The out-of-plane conductivity, \(\kappa_{zz}\), is shown in panel c. Maximum off-diagonal terms of the conductivity tensor are four orders of magnitude smaller than diagonal terms (\(\kappa_{yz}\sim \kappa_{xz} \sim \kappa_{xy} \sim1.5\times 10^{-4}\) W/m-K). They are not shown as they will have a negligible impact on the heat flow.}
    \label{fig:conductivity-map}
\end{figure}

As a demonstration of this workflow, we simulate the first 11 levels of metal (6 via and 5 line levels), the surrounding dielectric layers, and a 2~\micron{} field oxide capping layer in a 100$\times$100~\micron{} section from middle of a test die. 

\Fig{}~\ref{fig:conductivity-map} shows the spatial variation of the three diagonal components of the thermal conductivity tensor in the BEOL. The maximum off-diagonal terms of the conductivity tensor are \(\sim 1.5\times 10^{-4}\) W/m-K, or, four orders of magnitude smaller than the diagonal terms. These components will have negligible impact on the overall heat transfer.

One way to interpret this is that the eigenvectors of the conductivity tensor are highly aligned with the chosen coordinate system. This is not surprising as the metal components in the BEOL are axis-aligned. We expect that we may see some deviation in the real etched structures that may exhibit geometric deviations from the design files.

The overall structure of each of the three components is similar, with a tee-like structure with low conductivity. This non-uniformity is a reflection of the variations in the local density and overall structure of the as-drawn interconnect structure read from the GDSII file and is two orders of magnitude larger than the lines and vias in the RVEs.

Comparing the two cross-plane conductivities (\Fig{}~\ref{fig:conductivity-map}a and \Fig{}~\ref{fig:conductivity-map}b), we see that \(\kappa_{xx}\) is smaller than \(\kappa_{yy}\). The out-of-plane conductivity is significantly smaller than either of the in-plane conductivities. As can be seen in \Fig~\ref{fig:rve-exemplar}, this die is typical in that there are banks of lines running in the $x$ or $y$ direction on alternate line levels, but the cross sectional area fraction in the $z$ direction represented by vias is significantly less. With two orders of magnitude difference in the thermal conductivities of copper and dielectric, heat transfers mainly along the interconnect structure, most of which is in-plane.

These results are unique for the particular 100 by 100 \micron{} section of our test vehicle. We would expect that other locations and other test vehicles will have different relationships between the conductivities emphasizing the need for the type of automated workflow described in this work.

\begin{figure}
    \centering
    \includegraphics[width=\linewidth]{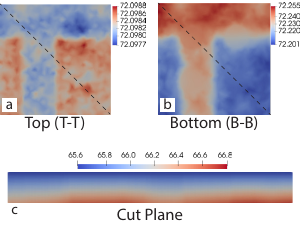}
    \caption{Temperature fields for the uniform case (uniform heat flux of 0.1256~W/mm\textsuperscript{2}) on (a) the top surface (T-T) and (b)~the bottom surface (B-B). Dotted lines in (a) and (b) show the location of a cut plane (c) that extends from the bottom (B-B) to the top (T-T) of the BEOL layer. Note that this cut plane has been stretched vertically to better show detail. Temperature fields show spatial variation mirroring the distribution of the thermal conductivity.}
    \label{fig:temperature-uniform}
\end{figure}

The temperature field for the uniform case, with uniform heat flux is shown in \Fig{}~\ref{fig:temperature-uniform}. The temperature fields closely match the structure of the thermal conductivities as seen in \Fig{}~\ref{fig:conductivity-map}. At the top of the BEOL (T-T, as defined in \Fig{}~\ref{fig:model-chip}), the temperatures are lower where the conductivity is lower, less heat having been transferred across the BEOL. On the bottom of the BEOL (B-B), the temperatures are higher where the thermal conductivities are lower, with more conductive pathways across the BEOL. It must be noted that the total variation due to this change in thermal conductivity is a fraction of a degree for the chosen parameters.

\begin{figure}
    \centering
    \includegraphics[width=\linewidth]{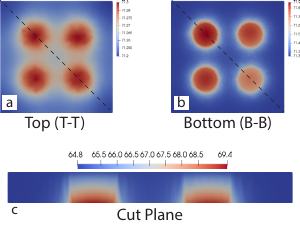}
    \caption{Temperature fields for the non-uniform case (heat flux of 1.0~W/mm\textsuperscript{2} on 4 contact patches) on (a) the top surface (T-T) and (b)~the bottom surface (B-B). Dotted lines in (a) and (b) show the location of a cut plane (c) that extends from the bottom (B-B) to the top (T-T) of the BEOL layer. Note that this cut plane has been stretched vertically to better show detail. Temperature fields show spatial variation mostly mirroring the non-uniformity of the heat flux.}
    \label{fig:temperature-patches}
\end{figure}

The non-uniform case, representing a heat load applied via microbumps is shown in \Fig{}~\ref{fig:temperature-patches}. On Bottom Surface, shown in \Fig{}~\ref{fig:temperature-patches}b, the temperature follows a similar trend to the uniform case where the temperatures are higher in regions of low thermal conductivity. This is seen on the bottom surface as the lower right microbump has the minimum temperature of the loading patch and the top left microbump has the maximum temperature of all the loading patches. Again, the total variation is not large for the chosen loading parameters, being approximately a third of a centigrade degree.

Interestingly, the top surface---\Fig{}~\ref{fig:temperature-patches}a---does not show a noticeable impact of the heterogeneity of the thermal conductivity. The scale of this response indicates that the in-plane diffusion is enough to mask the heterogeneity and that the size and pitch of the microbumps that provide the heat loading are a critical modeling parameter that cannot be ignored. The choice of these will determine how important the heterogeneity of the heat flux is on the top side of the BEOL.

\section{Conclusions and Limitations}
In this paper, we have demonstrated an automatic workflow for constructing spatially varying effective properties for the BEOL. Due to the linearity of the heat transfer problem, these properties can be computed a priori for a given macroscale mesh allowing the procedure to be used with industry standard finite element solvers such as ANSYS. In this work we use the MuMFiM multiscale finite element solver.

Construction of the BEOL material properties requires construction of 3D RVEs on which the homogenization procedure takes place. RVE construction required the development of custom modeling algorithms that can convert GDSII and OASIS~\cite{OASIS} files into 3D, Parasolid format, CAD models that can be automatically meshed.

Based on the homogenized conductivity values, we show the integration of the BEOL into a macroscale model that represents half of a chiplet stack. On the chip scale, we are able to utilize highly graded meshes with large elements away from the thin BEOL and FEOL layers.

The results indicate that the loading conditions will interact with the heterogeneity of the conductivity, and thus accurate modeling of the spatial variation of the thermal loading will be critical to understand the impact of BEOL heterogeneity. With heat flux loading that models microbumps, we only saw an impact of the variation in the conductivity on the bottom of the BEOL, but not on the top.

The automated nature of our workflow opens new avenues of exploration of chip-scale thermal design. This initial demonstration shows the effectiveness of the methodology, but many questions and potential limitations still remain to be investigated.

One of the major limitations of the current multiscale formulation is that it requires strong scale separation, or that the RVE does not feel the effect of varying spatial gradients as described in section \ref{sec:multiscale-model-formulation}. That is, the macroscopic temperature gradient is assumed to be a constant across the RVE (RVE contains fluctuations around macroscale gradient). This requirement sets up a competition of length scales between the size of the macroscale element that is wanted for accuracy, and the size of the RVE which should be an order of magnitude smaller than the macroscale element. In future works, we plan to carefully study this requirement and construct a set of best practices for RVE sizes in chip modeling.

\bibliography{references}
\bibliographystyle{IEEEtran}

\end{document}